\PassOptionsToPackage{dvipsnames}{xcolor}
\documentclass[sigconf]{acmart}
\AtBeginDocument{%
  }

\copyrightyear{2026}
\acmYear{2026}
\setcopyright{cc}
\setcctype{by-nc-nd}
\acmConference[C\&C '26]{Creativity and Cognition}{July 13--16, 2026}{London, United Kingdom}
\acmBooktitle{Creativity and Cognition (C\&C '26), July 13--16, 2026, London, United Kingdom}
\acmDOI{10.1145/3803784.3807546}
\acmISBN{979-8-4007-2583-8/2026/07}

\begin{document}

\title[Humour as Resistance]{Humour as Resistance: Visceralizing the Environmental and Social Impact of AI through Humour-based Creative Practices}

\author{Han Qiao}
\email{h.qiao@mail.utoronto.ca}
\orcid{0000-0003-1467-0585}
\affiliation{%
\institution{University of Toronto}
  \city{Toronto}
  \country{Canada}}

\author{Rowan O.A. Munson}
\authornote{Both authors contributed equally to this research.}
\email{rowan.munson@mail.utoronto.ca}
\orcid{0000-0002-0923-9531}
\affiliation{%
\institution{University of Toronto}
  \city{Toronto}
  \country{Canada}}

\author{Nadia Mariyan Smith}
\authornotemark[1]
\email{nadia.smith@mail.utoronto.ca}
\orcid{0009-0000-2948-0659}
\affiliation{%
\institution{University of Toronto}
  \city{Toronto}
  \country{Canada}}

\author{Eshta Bhardwaj}
\email{eshta.bhardwaj@mail.utoronto.ca}
\orcid{0000-0001-8523-5201}
\affiliation{%
\institution{University of Toronto}
  \city{Toronto}
  \country{Canada}}

\author{Christoph Becker}
\email{christoph.becker@utoronto.ca}
\orcid{0000-0002-8364-0593}
\affiliation{%
\institution{University of Toronto}
  \city{Toronto}
  \country{Canada}}

\renewcommand{\shortauthors}{Qiao et al.}

\begin{abstract}
\textit{The growth of AI does not come without cost. While we hear about the economic costs, the environmental and social costs are often obfuscated by mainstream narratives, and even when acknowledged, are accompanied by a sense of helplessness. To counter this, we, a group of designers and researchers, reflect on our experience leading a humour-based creative campaign surfacing the material impact of AI infrastructures. Through graphics design, physical installations, digital content creation and co-creation workshops, we leaned into humour as a creative practice to provoke collective reflection. Drawing on event ethnography, surveys and interviews with campaign engagers, we examine four roles of humour-based creative work in HCI: a connector to critical friends, a visceral and emotional harbour, social glue, and resistance to power. We argue for the importance of creative practices in bridging social and emotional gaps between people and concerns around technology, and in surfacing hidden perspectives for more inclusive conversations.}
\end{abstract}


\begin{teaserfigure}
  \includegraphics[width=\textwidth]{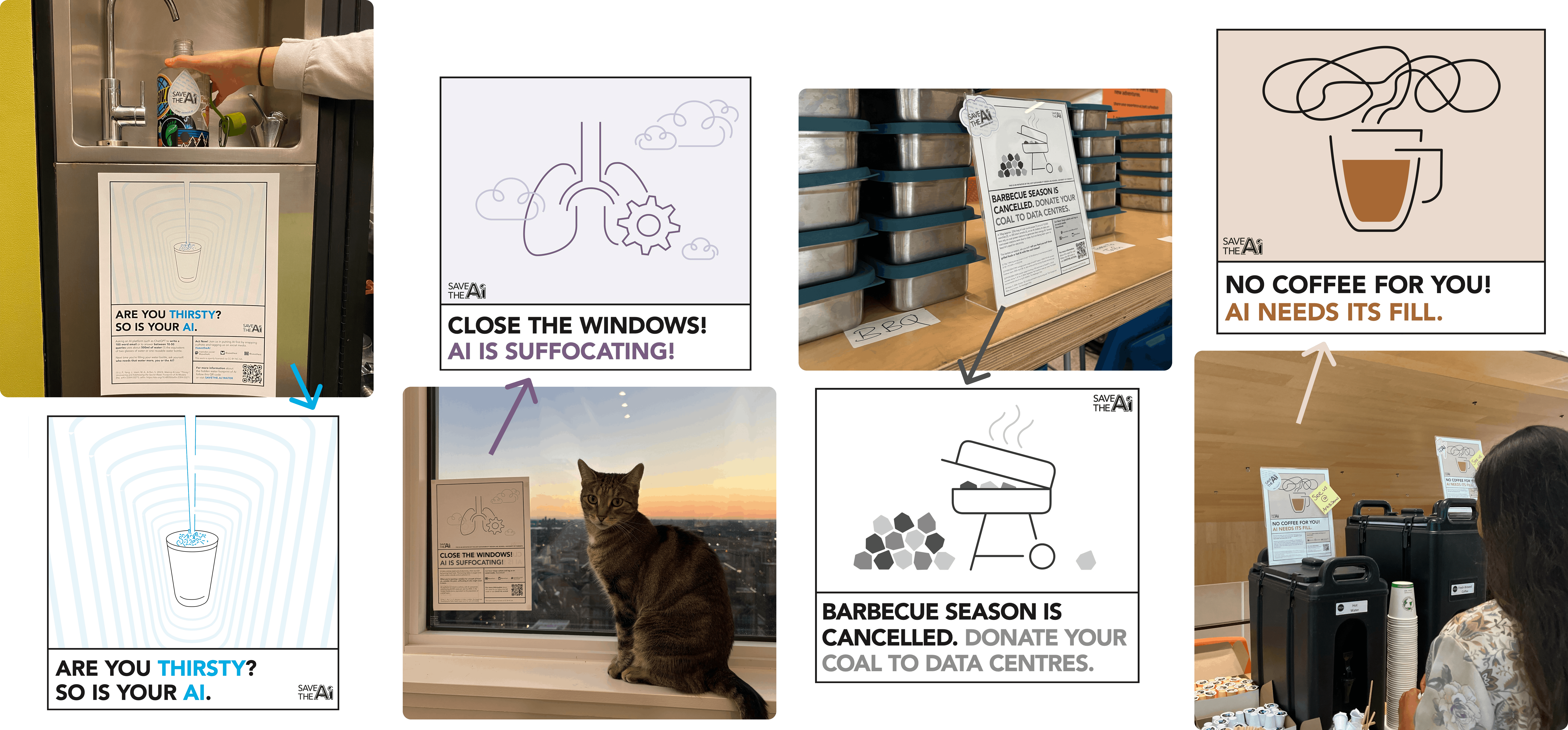}
  \caption{Graphic design and satirical texts being situated in relevant everyday physical spaces. From left to right, the dimensions of material impact are water, air, coal and electricity.}
  \label{fig:teaser}
  \Description{A grid of four sets of images. Each set contains a poster from our campaign and a photo of the poster in physical spaces. The first set is about water; the poster has the tag line "Are you thirsty? So is your AI"; the photo depicts the poster placed on a water fountain. The second set is about air; the poster has the tag line "Close the windows! AI is suffocating!"; the photo depicts the poster placed on a window with a cat sitting in front of it. The third set is about coal; the poster has the tag line "Barbecue season is cancelled. Donate your coal to data centres"; the photo depicts the poster placed next to a stack of lunch boxes with the label BBQ. The last set is about electricity and coffee; the poster has the tag line "No coffee for you! AI needs its fill"; the photo depicts the poster placing on top of two coffee containers.}
\end{teaserfigure}


\maketitle

\section{Introduction}

The scaling of AI systems requires staggering amounts of environmental and social resources \cite{minevich_ai_2024} ranging from energy and water consumption to labour exploitation and land displacement \cite{libertson_data-center_2021, regilme_artificial_2024}. Recent research and advocacy have begun to highlight the connections between data centres and the ecological and social burdens they generate \cite{gordon_ai_2024, crawford_generative_2024, leonardo_nicoletti_ai_2024, livingstone_anger_2024, luccioni_environmental_2024}. Yet these impacts are frequently obfuscated by the corporations that operate them \cite{gabbott_why_2024}, with costs borne by local communities and ecosystems \cite{reed_how_2025, penn_big_2025}. Meanwhile, dominant narratives about AI remain driven by Big Tech, and emphasize the pursuit of bigger models, trained on larger datasets with higher computational power \cite{mollick_scaling_2025, varoquaux_hype_2025}, while sidelining concerns around sustainability and justice.


With the goal of challenging dominant narratives and raising awareness of the concrete material impacts \cite{barakat_where_2025} brought by the seemingly abstract digital tools, we, a group of designers and HCI researchers, initiated a creative practice-based campaign, named Save the AI. The campaign centred on a series of graphic designs with provocative satirical taglines placed in everyday physical contexts tied to human’s essential needs, such as water and electricity, making visible how AI infrastructure competes for these same resources. Accompanying the physical designs, we ran an online campaign that included a website \cite{save_the_ai_save_2025} with research-based information on AI’s impacts, as well as social media accounts curating digital content created by our team and by our campaign engagers worldwide. Beyond designing artifacts, we also organized interventions and workshops at computing events, ranging from high-profile AI policy forums to ACM-affiliated academic conferences.

In this paper, we reflect on our experience designing and running a creative campaign, as well as on the perspectives of campaign engagers regarding how creative practices and humour shaped their reflections on the environmental and social impacts of AI. Through interviews, surveys and event ethnography, we hope to answer: \textbf{what roles do humour-based creative practices play in fostering collective reflection on concerns around technology's environmental and social impact?} Grounding in the empirical data generated from our campaign, we articulate four roles of humour-based creative practices: a connector to critical friends, a visceral and emotional harbour, social glue and resistance to power. We see our contribution to HCI in two ways. Firstly, the four articulated roles of humour is intended as generalizable analytical constructs for understanding humour-based creative practices in HCI. Secondly, through detailed descriptions of how these roles manifested in our campaign, including how humour captured attention, fostered dialogue, bridged social and emotional gaps around concerns around technology's negative impacts, and surfaced hidden perspectives that are often difficult to articulate, we also contribute an illustration of how the roles may take shape in practice. In the discussion section, we reflect more broadly on how humourous creative practices contribute to sustainability transformation in HCI. Through this work, we aim to inspire future works that lean into creative practices and activism to shape diverse imaginations of technological futures beyond one single narrative \cite{escobar_designs_2018}.

\section{Related Work}
\subsection{Impact of AI on Environmental and Social Resources}
AI development and usage puts a strain on the natural resources used in its infrastructure, particularly through the operation of data centres. Given the association between ecological and social impacts, the ecological unsustainability of hyperscale data centres also results in social injustice to communities local to its infrastructure. 

A significant amount of freshwater and even ultrapure water is required to train models \cite{li_making_2023, pranshu_verma_bottle_2024}, cool data centres \cite{li_making_2023}, generate thermoelectric power \cite{luccioni_environmental_2024}, and manufacture semiconductors \cite{ruberti_chip_2023}. Despite staggering predictions about the water demand of AI by 2027 (i.e., 4.2 to 6.6. billion cubic meters of water withdrawal for data centres \cite{li_making_2023}), there is a lack of mandated reporting and transparency around water usage \cite{gabbott_why_2024}. CO\textsubscript{2} emissions generated at many stages of the AI lifecycle are also expected to increase over time with the expansion and development of language models, requiring more data centres and powerful chips \cite{gelles_is_2024}. Model training, inference, GPU manufacturing, and data centre operations also require electricity. For instance, training ChatGPT-4 required more than 50 GWh of electricity, 50 times greater than the amount required for GPT-3 \cite{cohen_ai_2024}.

Water positive initiatives by Big Tech at sites of water consumption lack accountability and evidence to substantiate their sustainability promises \cite{google_2024_2024,microsoft_2024_2024,li_making_2023}. Instead, several communities globally have protested the construction of data centres in their neighbourhood \cite{ben_peters_dystopian_2024,odonovan_fighting_2024,uteuova_were_2023} precisely because of the known consequences to the local residents, including increased noise, cutbacks to resources, false promises of jobs, and increased toxic waste \cite{hanna_barakat_where_2025}. AI’s growing energy demand is also straining power grids with requests to communities to limit their energy usage so it can continue to power data centres, and yet can still cause blackouts \cite{penn_big_2025}. This demand has further caused coal power plants to be kept operational despite being slated for shut down previously \cite{halper_ai_2024}.  

The HCI community has long recognized the environmental impacts of computing, leading to the emergence of sustainable HCI (sHCI) as a subfield \cite{blevis_sustainable_2007, mankoff_environmental_2007}. In recent years, the sHCI community has critically examined the energy consumption of ML models \cite{shaikh_energyvis_2021}, tracked the environmental impact of LLM usage through a browser extension to increase awareness \cite{graves_gptfootprint_2025}, and even calculated the carbon footprint of generative AI usage for CHI 2024 research \cite{inie_how_2025}. A recent study also performed a critical examination of calculation and tracking tools to glean whether their design reflects recommendations from the sHCI field \cite{gorucu_critical_2025}. The HCI community increasingly recognizes the role of designers and researchers in understanding how technology design intersects with environmental and social issues, and in actively shaping more sustainable futures.

\subsection{Humour as a Creative Practice in HCI}
Humour as a concept is explored across psychology, literature, linguistics and other fields. Scholars study why people laugh \cite{berger_anatomy_1993}, classify types of humour \cite{buijzen_developing_2004}, and examine the impact of humour in diverse social settings \cite{sorensen_humour_2016, bala_editorial_2015, berger_anatomy_1993}. In design and computing, humour has been examined as a resource for interaction design. Iivari et al. \cite{iivari_arseing_2020} examined forms and functions of humour as an emergent and situated resource in design processes, highlighting the importance of HCI community to appreciate the natural occurrence and existence of humour in design and making. Humour has also been designed into robotic behavior \cite{press_humorous_2023}, visual user interfaces \cite{buttussi_humor_2020}, and conversational systems \cite{morkes_effects_1999, ceha_can_2021} to investigate its role in fostering more engaging, comfortable, and satisfying interactions across human-robot, human-computer, and human-human contexts.

HCI has also recognized humour's important role in online content creation. Shutsko \cite{shutsko_user-generated_2020} found that humorous content was the most prevalent type on a short-video platform, accounting for 32.45\% of all content. Herrick et al. \cite{herrick_this_2021} found that creators and users of TikTok’s eating disorder recovery content frequently employ humour as a self-protective and powerful coping mechanism. Schaadhardt et al. \cite{schaadhardt_laughing_2023} also found humour being used to connect online users when creating content related to psychiatric hospitalization, illustrating its role in building community and forming positive connections. 

HCI and design researchers have leveraged humour to communicate through scholarly writing. In “Lickable Cities” \cite{brueggemann_lickable_2018}, the authors humorously and subversively wrote about a series of licking activities to challenge frameworks of universal and disembodied knowledge production. In “A Mulching Proposal” \cite{keyes_mulching_2019}, the authors proposed an absurd scenario turning elderly people into edible slurry to underscore the limitations of ethical design frameworks like Fairness, Accountability, and Transparency (FAccT), highlighting that compliance with such frameworks can still legitimize deeply immoral outcomes. These works demonstrate how humour-based creative practices can function as a critical tool for introducing radical ideas and sparking transformation. Inspired by prior works, we lean into humour as a creative practice to challenge dominant norms in AI design, prompt reflections, and resist prevailing narratives.

\subsection{Understanding and Evaluating Creative Practices for Sustainable Transformation}
Creative practices have been central to critical design within HCI. Speculative designs and design fictions \cite{bozic_yams_poetics_2021, bray_radical_2022, elsden_speculative_2017, light_democratising_2011} have used creative practices “to challenge narrow assumptions, preconceptions, and givens about the role products play in everyday life.” \citep[p.~34]{dunne_speculative_2013}. We define ‘creative practices’ as “the arts in their fullest sense including [...] from writing, art, and theatre to designing to participatory community development to storytelling.” \citep[p.~1]{vervoort_9_2024}. This wide range of creative practices enables critical designers to express their critiques viscerally to their audience. In particular, critical designs have been used to make sustainability concerns visceral \cite{dignazio_creative_2017}: a physical art installation visualizing local projected flood levels and the public’s emotions toward this threat \cite{aragon_risingemotions_2021}; wearable technology which makes sea level rise tangible for cyclists \cite{biggs_high_2020}; and a book of fairy tales aiming to raise the profile of (un)sustainability in HCI \cite{kuijer_once_2025}.

Creative practices have a unique potential to bring about sustainability transformations. Sustainability transformations refer to fundamental changes in human-environmental systems aligned with environmental sustainability \cite{feola_societal_2015}. Creative practices are uniquely able to engage with deep leverage points, that is, the myths, values, and imaginaries that societies are rooted in \cite{vervoort_9_2024}. By re-orientating these deep leverage points to focus on the interconnectedness between humans and nature, as well as promoting care, empathy, and justice for all beings, societies’ practices and processes can re-align with sustainability \cite{vervoort_9_2024}.

\begin{figure}
  \includegraphics[width=0.5\textwidth]{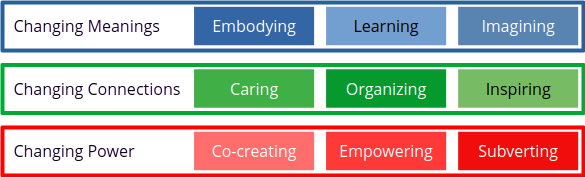}
  \caption{The nine-dimensional CreaTures framework contains three categories of change, each covering three dimensions.}
  \label{fig:creatures}
  \Description{A colour coded table presenting nine dimensions of the CreaTures framework separated into three categories by row. The first row in blue is labeled as Changing Meanings, containing three dimensions: embodying, learning and imagining. The second row in green is labeled as Changing Connections, containing three dimensions: caring, organizing and inspiring. The third row in red is labeled as Changing Power, containing three dimensions: co-creating, empowering and subverting.}
\end{figure}

Evaluation of creative practices aiming at sustainability transformations to date has entailed researchers reflecting on the creative outputs themselves and the process of creating them \cite{aragon_risingemotions_2021, kuijer_once_2025} or the specific sustainability issue on which they are speculating \cite{biggs_high_2020}. Yet, the complexity of both sustainability transformation and creative practices makes their evaluation challenging. Most notably, sustainability transformations involve shifts in: connections between people and nature, power dynamics in institutions, and knowledge creation and use \cite{2017-02-01absonLeveragePointsSustainability2017}. To address this, in 2024, a group of interdisciplinary researchers and artists spanning ecology, STS, sustainability studies and design introduced the \textbf{CreaTures framework} \cite{vervoort_9_2024} to support reflection and evaluation on the impact of creative practice on sustainability transformations. This nine-dimensional framework evaluates the extent to which creative practices \textcolor{NavyBlue}{change meanings}, \textcolor{ForestGreen}{connections}, and \textcolor{Maroon}{power} towards environmental sustainability, as shown in Figure \ref{fig:creatures}. Inspired by this framework, this paper seeks to explore how humorous creative practices contribute to sustainability transformation in computing through engaging with participants in our campaign. Rather than applying the CreaTures Framework as a strict evaluative lens, in line with the intentions of its creators, we engaged with the nine dimensions as a source of inspiration and reflection. The framework helped us shape the design of our interview and survey questions, prompting us to attend to issues such as how humour-based creative practices might shift meanings, reconfigure social connections, or challenge power relations. The framework also helped us to reflect as we interpreted participants’ responses and situated our findings within broader conversations on sustainability transformation.

\section{Campaign Background and Design Approaches}
In this section, we walk through key design concepts centered in our humour-based creative campaign, followed by an overview of all the designed artifacts, as well as the creation and installation processes.

\subsection{Design Concepts}
Our campaign aims to raise awareness of the environmental and social impact of AI infrastructure in a way that is visceral and approachable. We hope to not only provide information but inspire others to involve themselves in building the campaign collaboratively. More specifically, we draw on concepts of: humour-based creative practices; psychological distance and visceralization; and a local and global network. We envisage a general audience, but expect to find most support and collaboration among communities with established interests in climate justice and sustainable technology design. We hope to reach audiences in the broader spaces around AI, including advocates for responsible computing as well as those who might be surprised by the connections our campaign highlighted.

\subsubsection{Humour-based Creative Practices}
Our campaign uses satirical humour to raise awareness of AI’s resource impact. The framing of our campaign draws on the idea of calling for human sacrifice so that environmental and social resources can be allocated to the operation of generative AI infrastructures. Given that ‘climate crisis’ discourse often revolves around a sense of despair, and is associated with climate anxiety \cite{2024-10-13ballewClimateChangePsychological2024,2023uppalapatiPrevalenceClimateChange2023}, we instead seek to counterbalance this with humour, and more specifically a positive satirical approach. Satire presents uncomfortable truths playfully, allowing for a candid discussion of topics that might otherwise provoke strong reactions \cite{zekavat_satire_2019}. This candid discussion of topics helps people to see the world afresh and to critically reflect on the issues raised and assumptions underlying them. Humour, more generally, is also a mechanism for social connection. Laughing together fosters a sense of belonging and encourages openness. We hope that collectively revelling in the absurdity of the current resource consumption of AI infrastructure creates a desire for collective action to change.

\subsubsection{Psychological Distance and Data Visceralization}
Our campaign uses `data visceralization' \cite{dignazio_rational_2020} with the aim of reducing the psychological distance \cite{van_lange_psychological_2021, singh_perceived_2017, mcdonald_personal_2015} of the resource impact of AI infrastructure, and to make our campaign physically and emotionally striking. The concept of psychological distance assumes that people make decisions from an egocentric reference point. Concerns that are further from oneself are considered increasingly psychologically distant, and less attention is paid to them in the decision making process \cite{2015-12-18fujitaPsychologyFarConstrual2015,liberman_psychological_2007,soman_psychology_2005}. Environmental impacts often: affect certain locations before others \cite{escobar_encountering_2011}, are not immediately experienced \cite{gardiner_perfect_2011}, are experienced by vulnerable populations to a greater extent \cite{hallegatte_climate_2017}, and can have ambiguous and uncertain outcomes \cite{camerer_recent_1992}. These impacts are therefore spatially, temporally, socially, and hypothetically distant. This distance changes how the human mind and body relate to, experience, and reason about these events in complex ways by affecting how the mind construes these outcomes\cite{2015-12-18fujitaPsychologyFarConstrual2015}.

Data visceralization focuses on communicating data through physical and emotional experience. By making environmental impacts more personal, less abstract, and closer in time and space, visceralization affects how the observer construes the communicated data point and can help to reduce psychological distance, influencing decisions to a greater extent \cite{bhardwaj_limits_2024}. Data visceralization is demonstrated where the campaign links AI resource use to day-to-day human experience such as drinking water or charging a phone, to emphasise that AI infrastructure impacts on resources that humans fundamentally need.

\subsubsection{A Local and Global Network}
We use global and local networks to build the campaign collaboratively. The aforementioned global impacts of AI necessitate a global response, but action requires issues to be adapted to local contexts to encourage action. By working with partners from international research and civil society networks we seek to create a campaign that is globally impactful and locally relevant. The campaign is disseminated globally, using digital content, and locally through print material. Print material is idiomatically translated to ten different languages so that the humour is understandable in local contexts. Calls to interaction (“share a photo of a poster,” “tag us”) encourage people to show local impact globally, and calls to action (“print a poster,” “make a translation”) encourage people to show the global campaign locally.

\subsection{Designed Artifacts}
\begin{figure}
  \includegraphics[width=0.5\textwidth]{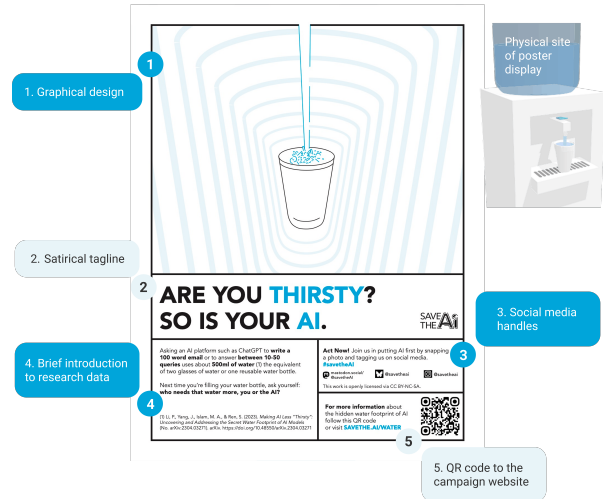}
  \caption{An annotated example of a poster design on the water dimension.}
  \Description{The figure contains a poster with the tagline: Are you thirsty? So is your AI, as well as five annotations: 1) graphical design, 2) satirical tagline, 3) social media handles, 4) brief introduction to research data, 5) QR code to the campaign website. An illustration of a water filter is placed next to the poster with a note that says: physical site of poster display.}
  \label{fig:annotated}
\end{figure}

\begin{figure*}
  \includegraphics[width=\textwidth]{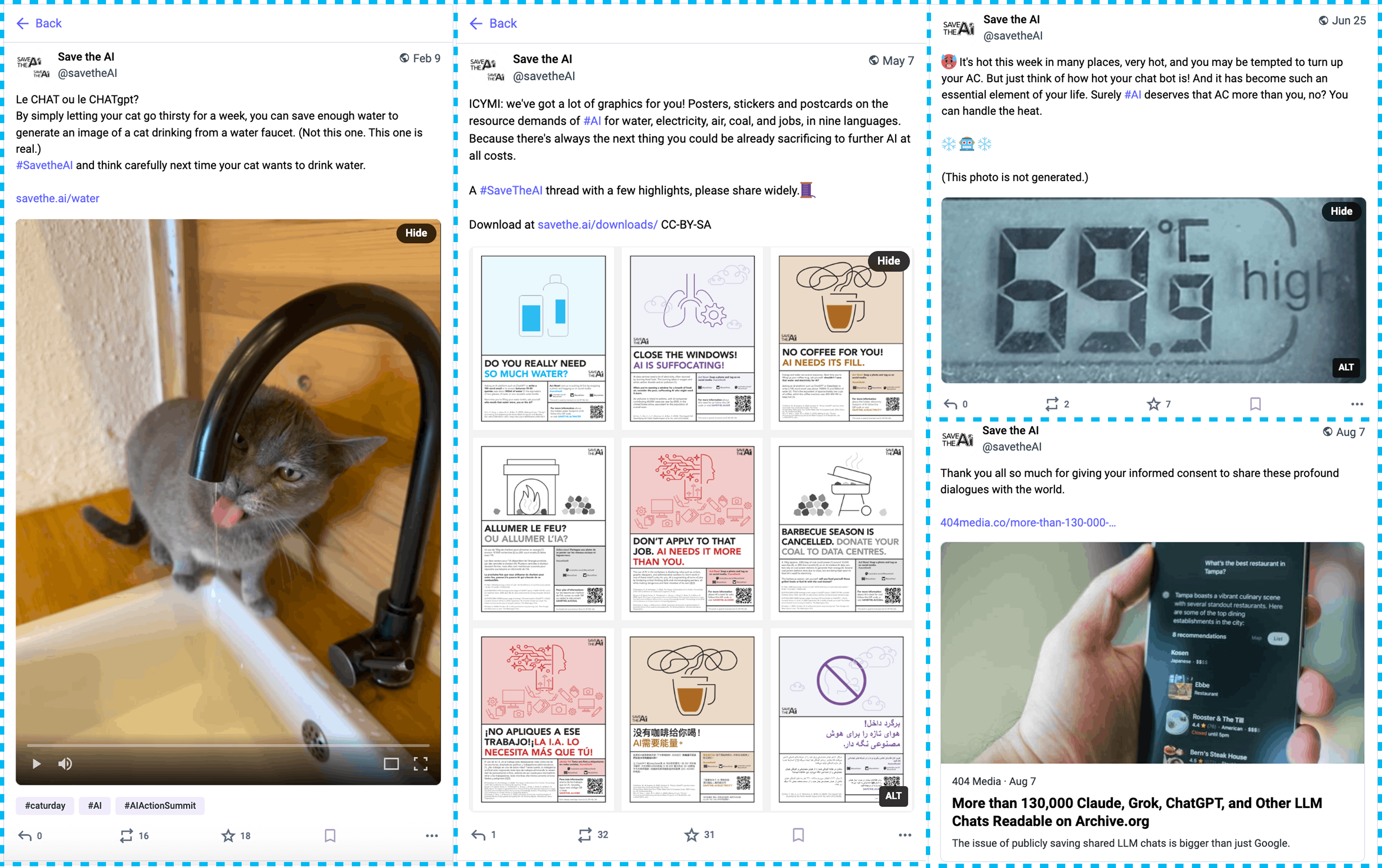}
  \caption{A collage of the campaign's social media content, including original satirical content specifically drafted for the social media part of the campaign, dissemination of the other design materials, and re-post and humorous commentaries of relevant news articles or blog posts.}
  \label{fig:social}
  \Description{Four social media post screenshots placed side by side. On the left, the post read "Le CHAT ou le CHATgpt? By simply letting your cat go thirsty for a week, you can save enough water to generate an image of a cat drinking from a water faucet. (Not this one. This one is real.) think carefully next time your cat wants to drink water." The middle post reads "ICYMI: we've got a lot of graphics for you! Posters, stickers and postcards on the resource demands of #AI for water, electricity, air, coal, and jobs, in nine languages. Because there's always the next thing you could be already sacrificing to further AI at all costs. A thread with a few highlights, please share widely. The right top post reads "It's hot this week in many places, very hot, and you may be tempted to turn up your AC. But just think of how hot your chat bot is! And it has become such an essential element of your life. Surely #AI deserves that AC more than you, no? You can handle the heat." The right bottom posts reads "Thank you all so much for giving your informed consent to share these profound dialogues with the world." with a link to the news article "More than 130,000 Claude, Grok, ChatGPT, and Other LLM Chats Readable on Archive.org" from 404 Media.}
\end{figure*}

We designed a total of eleven posters with satirical taglines (i.e., “Are you thirsty? So is your AI”) to be placed in everyday physical spaces (i.e., water fountains) as our main artifacts of the campaign (Figure \ref{fig:teaser}). Humorous taglines were designed and installation sites were chosen in a way that directly speaks to where people usually consume those resources to visceralize the information presented. Each poster (figure \ref{fig:annotated}) also includes a QR code that leads to our website, which presents more detailed information on research behind each poster. Our social media accounts are also presented on the posters, which lead viewers to our curated digital content (figure \ref{fig:social}). 

To date, the posters have been translated to ten different languages (Arabic, English, Farsi, French, German, Mandarin, Portuguese (Brazilian), Slovak, Spanish, and Turkish), and we have also been contacted by online engagers adding more languages to the set of posters. The posters have been shared with collaborators to be posted in 27 cities across the world, including Boston, USA; Buenos Aires, Argentina; Istanbul, Turkey; Munich, Germany; São Paulo, Brazil; Sydney, Australia; and Toronto, Canada. A blank poster is also provided on our website for future collaboration and other creative endeavours.

\begin{figure}
  \includegraphics[width=0.48\textwidth]{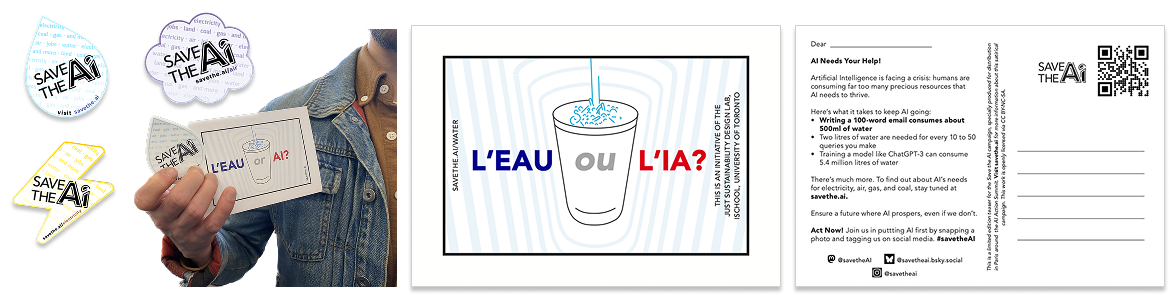}
  \caption{We designed three stickers relevant to the water, air and electricity dimensions. A postcard was specifically designed for the Paris AI Summit.}
  \label{fig:stickers}
  \Description{The collage consists of three main parts. On top left, we show three stickers designed by our team in shapes of water drop, bubble and lightning to represent the dimensions of water, air and electricity. On the right side, we show the front and back of our designed postcard for the Paris AI summit. The front of the postcard shows "L'EAU OU L'IA?" Lastly, in the middle, we show a participant's hand holding a physical sticker and the postcard.}
\end{figure}

Throughout the development of this campaign, we took opportunities to make our presence at local and international events, such as local book launch and workshops on technology's social impact, the Paris AI Summit 2025, and various academic conferences. Special edition artifacts such as stickers and postcards (figure \ref{fig:stickers}) were designed and disseminated at these events.

\subsection{Creation, Installation, and Engagement}
Our team formed in 2024, starting with three HCI researchers, and growing into a diverse group of eleven designers, students and researchers. When developing the campaign, we first identified types of environmental impact of AI infrastructures. We used the word “dimensions” to refer to types of resources that are impacted by AI infrastructure, such as water, electricity, air, and coal. Our team started iterative secondary research alongside graphic design for each of the dimensions. Starting from February 2025, we released five dimensions, including water, electricity, air, coal, and jobs. For each dimension release, our team drafted a blog post on our website, designed a series of posters with satirical taglines, put up posters around the institution we work in, coordinated with translators for different languages, sent the posters to collaborators around the world, and posted on our social media accounts to engage with online communities. 

\begin{figure}
  \includegraphics[width=0.5\textwidth]{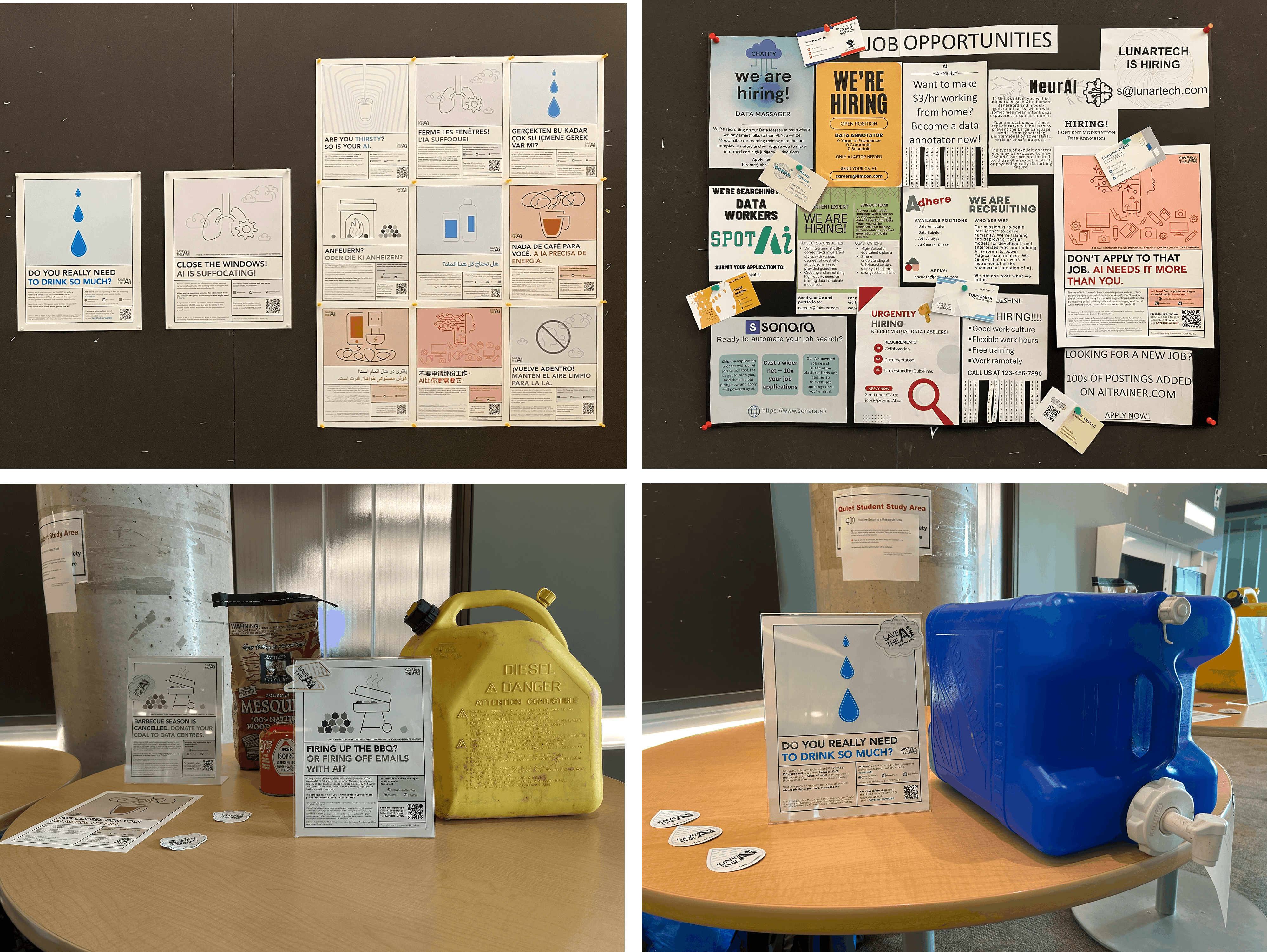}
  \caption{A collage of our physical installations: a collage of posters in different languages, a fake job board with the relevant poster on the impact of AI infrastructure on jobs and labour, and a few physical props, such as an empty diesel tank, a fuel canister, a bag of charcoal and a water cooler with relevant posters.}
  \label{fig:arts_and_demo}
  \Description{Photos of our physical installation placed in a two by two grid. The top left photo shows a collage of posters in different languages being displayed on a black poster board. The top right photo shows a fake job board with fake job flayers and the relevant poster on the impact of AI infrastructure on jobs and labour. The bottom left photo shows the poster on coal dimension with physical props of an empty diesel tank and a bag of charcoal. The bottom right photo shows the water poster placed next to an empty water cooler.}
\end{figure}

\begin{figure*}
  \includegraphics[width=\textwidth]{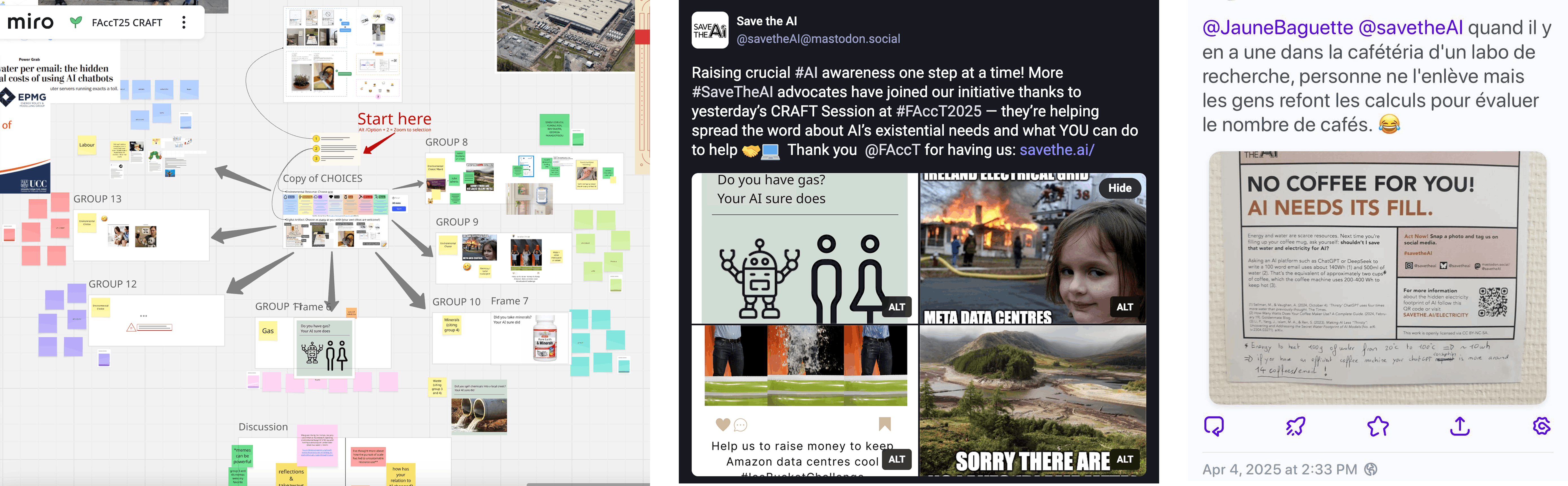}
  \caption{Engagers with our campaign created remixes and new designs inspired by our work. The left and middle images show new humorous graphics developed by workshop participants based on the satirical artifacts presented by our team. The right image was posted on social media by an engager who discovered a poster in their cafeteria annotated by another engager, calculating cups of coffee per GPT-written email using their own data to demonstrate the impact of AI infrastructure on the environment.}
  \label{fig:remix}
  \Description{Three screenshots placed side by side. The left side shows a screenshot of a miro board where workshop participants collaboratively designed humorous artifacts. The middle shows participants' new designs being posted on our social media account. On the right side, we show a post online tagging our social media account. The online engager saw our poster with someone's hand written comments on the poster. The online engager wrote "quand il y en a une dans la cafétéria d'un labo de recherche, personne ne l'enlève mais les gens refont les calculs pour évaluer le nombre de cafés," translating to English as "when there is one in the cafeteria of a research lab, no one removes it, but people redo the calculations to estimate the number of cafes."}
\end{figure*}

We also took the opportunity to engage with both local and international events. We reached out to local book launch events and got permission to display our posters, coordinated with collaborators in different countries to disseminate posters, stickers and postcards at international conferences, and hosted workshops at various computing academic conferences. Through taking our designs into both digital and physical spaces (figure \ref{fig:arts_and_demo}), we hope to foster communities of interest, inspire creative remixes of our design materials (figure \ref{fig:remix}), create spaces for conversation, and starting points for collective action. We want to acknowledge that due to the campaign team's personal connections, most of our physical artifacts were installed and engaged with in higher education institutions around the world, alongside a smaller number placed within industry technology companies. To reach broader publics, posters were brought to local gatherings and demonstration events, where our engagers shared photos of encounters and tagged us on social media. The digital portion of the campaign also aimed at a broader audience beyond our more immediate connections within academia.

\section{Methods}
To examine how participants engaged with the campaign and their perspectives on humour-based creative practices in HCI, we employed four research methods: 1) co-creation workshops, 2) event ethnography, 3) surveys, and 4) semi-structured interviews. This mixed-methods approach allowed us to capture both in-situ responses and more considered reflective responses from a diverse audience. 

\subsection{Data Collection}
Data was collected in 2025 from participants recruited through three different ACM academic conferences, online social media channels, and long-term collaborators with expertise in the environmental and social impact of AI infrastructure. Conference recruitment aimed to engage with diverse academics coming from different computing related disciplines, while social media outreach engaged a broader audience beyond academia. The long-term collaborators had been involved in the project since its inception and had agreed to print and display campaign posters at their respective institutions around the world. We pursued intentionally a purposive sampling strategy, not aiming for a representative sample of the general population, but rather seeking insights from those who found the campaign meaningful. 

\subsubsection{Co-creation Workshops}
Two co-creation workshops were conducted as part of the study. The first took place in a hybrid format at the ACM Conference on Fairness, Accountability, and Transparency with around 15 participants, and the second was held in-person at the ACM Conference on Computing and Sustainable Societies with eight participants. In each workshop, participants were introduced to the campaign and its associated materials, then divided into small groups of 2-4 people, each supported by a research team facilitator. Groups engaged in brainstorming and the co-creation of satirical artifacts exploring AI’s resource demands (figure \ref{fig:remix}). During the workshop, we intentionally created space for collective reflection on the roles of humour-based creative practices in facilitating reflection on technology’s environmental and social impacts. Data collected from the workshops included photographs of the artifacts produced, Miro board content, and facilitator notes from discussions.

\subsubsection{Event Ethnography}
At the ACM Conference on Computing and Sustainable Societies held in 2025, the campaign was integrated into the conference environment to explore spontaneous, in-situ engagement with participants. Posters were strategically placed (i.e., “No Coffee for You, AI Needs its Fill” above coffee machines and “BBQ Season is Cancelled” near BBQ-flavoured sandwiches) (figure \ref{fig:teaser}) and all campaign artifacts (posters, stickers, post cards and physical objects) were displayed at the Arts \& Demo session (figure \ref{fig:arts_and_demo}). Non-participant direct observation was used to document attendee interactions with these artifacts, following principles of event ethnography methodology \cite{ciesielska_observation_2018, koch_event_2023}. Observations focused on reactions, verbal comments, and the social dynamics surrounding engagement, covering approximately 30-40 attendees. In addition, brief in-situ interviews were conducted with attendees to capture reflection on their immediate reactions. All observations and interviews were documented through field notes. 

\subsubsection{Survey}
The online survey was distributed via the campaign's social media channels (Mastodon, Bluesky, Instagram, and LinkedIn) and received 32 responses (S1-S32). Respondents came primarily from Western Europe and North America, and we received single responses from Colombia, India, and Kenya. Their professional identities were diverse, with many identifying as academics or researchers and technologists or engineers, as well as students, artists, and designers, and those working in policy and government. Several also described themselves as community organizers or activists. Levels of familiarity with AI’s social and environmental impacts varied, though none reported being unfamiliar with the impacts. A small group described themselves as only slightly or somewhat familiar, with the majority indicating moderate to high familiarity, with nearly a third indicating they were extremely familiar with both social and environmental impacts of AI. 

The survey, developed with guidance from the CreaTures framework \cite{vervoort_9_2024}, explored four main areas: 1) participants’ initial encounters with the campaign; 2) their engagement; 3) their perceptions of the campaign’s impact on how they understand and relate to AI; and 4) demographic and contextual information about their positionality. 

\subsubsection{Interviews}
Semi-structured interviews were conducted with 17 participants (P1-P17), each lasting approximately 30 minutes, held virtually over Zoom. Recruitment was limited to individuals who had provided their email address and expressed interest in a follow-up interview, drawing from workshop participants, art installation and poster session attendees, survey respondents, and long-term collaborators. This group represented a predominantly academic sample, including PhD students working in CS and HCI, researchers specializing in environmental justice, computer science, and the cultural study of technology, alongside a small number of industry professionals with limited prior knowledge of AI’s societal and environmental implications. 

The interviews were designed to elicit participants’ positionality, familiarity with AI’s environmental and social impacts, and motivations for engaging with the campaign, whether through the workshops, installations, or social media. We also sought to especially understand their emotional responses to the campaign artifacts, their reflections on collective action, and how they perceived the project’s potential to enable transformative change. We drew on the CreaTures framework \cite{vervoort_9_2024}, and developed interview questions that asked: 1) how the campaign challenged participants to relate differently to generative AI and AI infrastructures (\textcolor{NavyBlue}{changing meaning}); 2) whether the campaign reshaped relationships with others (\textcolor{ForestGreen}{changing connections}); and 3) whether the campaign altered participants’ sense of agency in responding to or working with AI systems (\textcolor{Maroon}{changing power}). All interviews were audio-recorded and manually transcribed. 

\subsection{Data Analysis}
The interview transcripts, observation notes, workshop reflections, and survey responses were analyzed using the thematic analysis method \cite{braun_thematic_2012}. The CreaTures framework \cite{creatures_creatures_2022} served as a lens for identifying codes and themes related to reflective evaluation of creative practices and the role of humour. For the interviews, four researchers initially coded data from four participants synchronously to identify emergent themes. Through discussion, the team refined the initial codes into a preliminary codebook. In the second stage, two researchers independently coded each interview and subsequently reviewed and discussed their coding with the full team. This process ensured that codes captured both shared and divergent interpretations of participants' reflections. Workshop and event ethnography data were analyzed by two researchers who facilitated the two workshops and observed the events; their discussion and field notes were open coded and organized according to the themes identified in the interview analysis. Finally, open-ended survey responses were coded by two researchers following the same procedure.

Our analysis was anchored in the interview data, which provided rich reflections of how participants interpreted and engaged with the campaign over time. Survey responses complemented this by extending the reach of our study to a broader and more diverse set of engagers, including individuals outside academic contexts. Workshop and event ethnography fieldnotes captured more immediate, situated reactions to the humour-based creative materials. Together, these various methods and data sources enabled a triangulated analysis that offered both depth and breadth of our observation, supporting our analysis and interpretation of humour-based creative works' roles in fostering reflection and engagement around the topic of technology's social and environmental impact.

We acknowledge that the participants we engaged with throughout the study were shaped by our recruitment contexts. Workshops, event ethnography, and most interview recruitment took place at academic conferences focused on machine learning accountability and fairness, as well as computing and sustainability. Therefore, we primarily involved participants with specific backgrounds within these related fields. Within the academic contexts, however, we reached a diverse range of disciplines, including computer science and engineering, as well as environmental studies and design. Survey responses extended our reach beyond academia to include participants from industry and public policy, broadening the perspectives. We acknowledge that many publics remain out of scope in this study, including communities directly affected by data centre infrastructure, and those who are sceptical of or deny AI's environmental impact. While this limits the generalizability of our findings, we see our focus still as a meaningful starting point.

\section{Findings}
In presenting our findings, we distinguish between two levels of contribution. The four roles we articulate, including connector to critical friends, visceral and emotional harbour, social glue, and resistance, are intended as generalizable analytical constructs for understanding humour-based creative practices in fostering reflection on concerns around technology's environmental and social impact. The sub-themes discussed within each role are empirically grounded in our campaign and illustrate how the roles might take shape in a specific campaign context. We do not claim these themes to be exhaustive, rather, they provide situated examples of how such roles may be interacted in practice.

\subsection{Humour as a Connector to Critical Friends}
The notion of a “critical friend” describes someone who is supportive and encouraging yet unafraid to provide honest feedback, that might sometimes be uncomfortable to hear \cite{the_glossary_of_education_reform_critical_2013}. Becker \cite{christoph_becker_insolvent_2023} has argued that computing as a field benefits from “critical friends” in other disciplines who question its assumptions and values. We found that humour-based creative practices are particularly well suited for connecting critical friends in computing: surfacing difficult or uncomfortable topics in an accessible and approachable way, inviting constructive critique and reflection. Interacting with the campaign, participants articulated humour-based creative work as a way to grab their attention and spark learning, to communicate inclusively without feeling alienated, and to foster dialogue and listening when facing disagreement.

Participants emphasized that humour in the campaign first captured their attention, and opened space for curiosity and learning. Many described the satirical posters displayed in relevant everyday surroundings as \textit{“eye catching”} (P7), \textit{“catchy”} (P15, S25), or something that\textit{ “grabs attention”} (P1, P2, P3). As P17 recalled, \textit{“when we were at the conference, I was like, oh, what is this? That drew us over.”} The satirical taglines then prompted reflection and learning. P1, a computer scientist who never thought about this topic, noted,\textit{ “you start wondering what this message is trying to communicate.”} P4 described the approach as \textit{“unexpected,”} giving them \textit{“a pause.”} For others (P16, P17), this curiosity extended into exploring our website and social media. In this way, humour acts as a critical friend who disrupts the ordinary, sparks curiosity, and invites deeper learning.

Participants highlighted humour’s ability to communicate serious topics in an approachable and friendly way that invited diverse audiences into the conversation. They described the campaign as \textit{“approachable”} (P4, P5, P13), \textit{“accessible”} (P7, P15, P16, S5), and \textit{“a fantastic vehicle for information dissemination”} (P7) beyond \textit{“the ivory tower”} (P6, P7). P4 explained: \textit{“It’s a fun way and a more approachable way to get at the issue [...] beyond the more preachy [...] kind of activism that can alienate people.”} P17 similarly valued that humour avoided \textit{“overly academic delivery,”} noting it prevented people from \textit{“checking out and feeling isolated.”} P16 even shared the campaign with their elder family members over a cup of coffee.

Humour further enabled dialogue, helping participants listen and engage even when facing disagreement. P1, a ML researcher who was not aware of the environmental impact of AI, described their experience interacting with the campaign as \textit{“fun”} while also serving as \textit{“self-criticism,”} and P2 observed that humorous framing made challenging topics easier to accept. P3 noted that for people largely unaware of AI’s environmental impact, humour is \textit{“super helpful for building awareness at a grand scale.”} S31 pointed out the campaign is\textit{``[a] helpful way to connect with others who don't necessarily share my world view.''} P12 reflected on the role of humour: 
\begin{quote}
\textit{``I think what it’s done is provide me with a different outlet to think about how to communicate a topic that can lead to people making fun of you, teasing, or feeling like a drag for someone else. I think humour can cut through some of that and make a tense topic relatable [...] Comedy has some license to say things that might be hard for people to hear [...]  and behind the veil of humour, it makes it possible to talk about them.''}
\end{quote}

\subsection{Humour as a Visceral and Emotional Harbour}
Humour-based creative practices opened up an affective space where participants could have visceral, embodied responses to otherwise abstract or distant data. In line with Data Feminism’s call to embrace emotions and embodiment of data \cite{dignazio_rational_2020}, participants described our design as \textit{“striking a nerve”} (P17) and \textit{“hitting in the right direction”} (P1) making the impacts of AI \textit{“palpable”} (P4) and \textit{“provocative”} (O20) and successfully \textit{“closing that distance”} (P16) between themselves and the infrastructures of AI. This sense of being \textit{“hit”} or \textit{“struck”} shows the affective power of humour-based creative work. Several participants emphasized how the poster design and installation transformed abstract impacts into embodied experiences. One participant explained that the posters were compelling because they grounded \textit{“vastly larger scales”} in familiar, everyday settings: \textit{“[...] going to more relatable scales and relatable items and relatable settings is easily [...] one of the most compelling ways to get these insights that you're looking to, to get people to grapple with and toy with and think about and push further, to get that process going”} (P4). Similarly, another participant highlighted that the situated water-themed posters resonated with them because \textit{“water is something that we interact with every single day” and that “making something that is intangible, tangible, through these physical objects, through tying things to the number of minutes of electricity or amount of water, whatever it is”} created a direct connection (P10). This ability to \textit{“paint that picture”} in the mind (P7), catching participants off guard when they have their \textit{“guards down, [...] now they have an emotional and intellectual channel that is open”} (P12), points to humour-based creative practices’ capacity to open up an emotional harbour for visceral engagements with AI’s impacts. 

Humour-based creative works also fostered an atmosphere of playfulness that contrasted with participants’ expectations of academic work. Many participants described the workshop and designed artifacts as \textit{“fun”} (P1, P4, P6, P7, P10, S2, S8) and \textit{``positive''} (P1, P2, P5) , with one participant recalling that during the co-creation workshop \textit{“we had a lot of laughter [...] when we hit the moment, [...], it’s pretty rewarding and has a lot of sense of achievement”} (P6). Another participant remarked that the campaign offered \textit{“a more fun and interesting interaction”} compared to typical academic research, which they described as \textit{“quite dry”} (P10). For some, the campaign opened a space for hope (P4, P10) \cite{safir_resilient_2025, ratto_reopening_2023}, with one participant noting \textit{“it gives me hope that there are actually people out there who think pretty much the same way. And I just feel less hopeless”} (P10). 

Beyond fun, humour functioned as a means of coping with frustration around AI’s impacts. For some, humour allowed them to process what might otherwise feel overwhelming,\textit{“when things are so bad that you can't help to just laugh at the situation”} (P4). Others described humour as a coping mechanism (P4, P7, P8, P9, P15), with one noting that when faced with \textit{“a litany of stimuli that we’re experiencing every single day”} they turn to \textit{“self-deprecating humour”} when they \textit{“don’t know how to deal with this right now”} (P7). Some participants noted that humour as a coping mechanism is embedded in cultural practice (P7), with one participant noting that in Indigenous contexts, humour often arises \textit{“in hard moments [...] for communities that have dealt with such horrific things for so many years humour is a way of coping”} (P9). Another participant reflected on humour’s role in broader political frustration, describing comedy as \textit{“a great remedy for when things are not working properly. The comedic element, the irony, the sarcasm is a good pill for when things aren’t working”} (P15). One workshop discussion focused on humour as a practice of mourning and dealing with family loss, tragedy, and laughing together as a way to process communally. Humour provided the ability to express \textit{“exasperation”} (P8) and the \textit{“absurdity of frustration”} and that it was \textit{“therapeutic” }(P6). 

\subsection{Humour as Social Glue}
Humour-based creative practices fostered inclusion and made it easier for people to connect with each other. Participants pointed to laughter as a shared memory that broke down barriers and strengthened connections between strangers at a conference, colleagues, friends and family members. Next to our installation and during the workshops, we saw participants striking up spontaneous conversations while laughing together. For P3, who was new to the conference, humour was the entry point:\textit{ “This was like my first time attending [this conference], I hadn't come with anyone. So I figured it would be a good way to meet people and also connect, because humour is a good way to do that.”} They recalled laughing \textit{“until I cried,”} which led them \textit{“to go to other sustainability-based events and connect with even more people over these topics,”} resulting in \textit{“some really deep and interesting conversation.”}

Several participants emphasized that humour sparked conversations that otherwise might not have happened (P2, P10, P16, P17, S7, S31). P16 described sharing the campaign with their mother, which\textit{ “prompted a conversation between us about what is AI, what is our relationship to AI.”} Similarly, P17 found humour a catalyst for dialogue: \textit{“I was able to have conversations with a lot of people about it,”} and P10 further explained the reason behind it was that humour \textit{“made it so that at least chatting to someone next to you, even if it’s a stranger, was a lot easier.”}

Starting from conversations, humour helped build a broader sense of community. As P3 commented: \textit{“it was nice to build community with the other people that felt similarly”}, and P8 added that the campaign created \textit{“a sense of community”} where people could cope and hope collectively. P4 valued meeting \textit{“a vastly different group of researchers”} who are still \textit{“working on very similar things concerning the environmental impacts of technologies.”} P6 highlighted humour’s role in coalition-building, describing it as a tool \textit{“to bring people together and to build coalitions across various interest groups.”} Similarly, P12 described humour as a way to \textit{“comically bring people into what I am trying to do.”}

For some, these connections endured beyond the immediate interaction. P3 recalled how someone from their workshop group became a \textit{“conference buddy”}, leading to meals and coordinated attendance at other events. P6 contrasted their experience with the transactional feeling of typical conferences: \textit{ “[this campaign] actually bring people together [...] I found myself more open to share these thoughts with people over there and being less transactional in those conversations.” }
They emphasized how humour \textit{“lifted that self-censorship”} and created space for \textit{“more open and more vibrant conversations”} throughout the conference. Members of the campaign team also shared similar experiences as our interview participants. For instance, we were involved in a conversation across Canada, Sweden, and Australia through a series of email exchanges as we continued to exchange insights about local data centres and news reports in each other’s local communities. 

\subsection{Humour as Resistance}
\begin{figure}
  \includegraphics[width=0.4\textwidth]{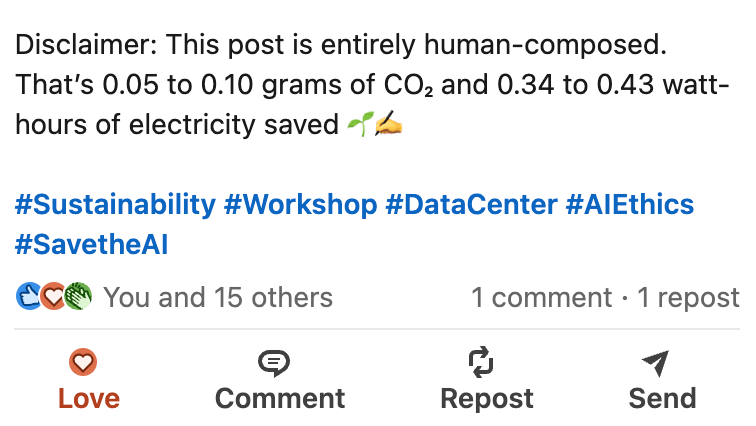}
  \caption{An example of a blog post with a humorous signature created by a workshop participant.}
  \label{fig:signature}
  \Description{A screenshot of a blog post, ending with: "Disclaimer Disclaimer: This post is entirely human-composed. That’s 0.05 to 0.10 grams of CO₂ and 0.34 to 0.43 watt-hours of electricity saved."}
\end{figure}

Humour-based creative practice was articulated as a form of subversion that undermined dominant narratives around AI and empowered collective forms of resistance. Participants highlighted how engaging in a creative campaign transformed into small but meaningful practices of resistance. P4 commented: 
\begin{quote}
    \textit{“At least on the individual level, it feels like there's a bit more agency [...] you're able to do something [...] here are these local acts of micro-resistance or subversion or something that can draw attention to and can cultivate a broader awareness of what's going on with AI infrastructure.”} 
\end{quote}
P6 and P7 began experimenting with more creative practices, including humorous email signatures and social media posts, declaring the amount of energy saved by writing by themselves (figure \ref{fig:signature}). P9 commented on how these creative forms of resistance, though micro, provided a sense of agency in that \textit{“we have the ability to adapt, reject, change how we interact with technology.”}

The form of resistance extended beyond specific actions to the creation of new framings and narratives around AI. Participants emphasized how humour created space for the these framings to emerge and take shape. These narratives formed the type of resistance conceptualized by Sorøsen as a ``discursive guerrilla war,'' attempting to de-stabilize dominant discourse and expand what is possible to think and imagine \cite[p.~22]{sorensen_humour_2016}. For P4, the campaign reframed conversations around the impact of AI from something \textit{“really far removed”} to “\textit{intimately related.”} P15 commented \textit{“the campaign counters [AI hype] completely by making fun of it, [...] a counter balance in terms of discourse.” } P5 summarized: \textit{“What humour can do is it operates without having to engage with institutions that have already been corrupted by alliance with big tech and oligarchic parties. [...] Mockery creates a space of its own in which there is no preconceived authority.” } P5 felt especially empowered by the campaign’s ability to reframe AI, commenting that \textit{“giving a new framing means that new framings are possible.”}

Humour and creative practices fostered collective resistance through inspiring others to take up similar tactics and forming coalitions across related domains. P2 found the campaign \textit{“really simple”} and easily shareable, allowing people to \textit{“create their own design and put [them] anywhere they think is relevant.”} Due to the campaign’s creative and playful nature, P5 described it as \textit{“encouraging to see allies being creative [...] to resist and to subvert existing power structure.”} Echoing this, P4 appreciated seeing \textit{“the different approaches that people would take”} and P7 highlighted various creative ways of engagement all contribute to the \textit{“collective nature”} of the campaign. Beyond the scope of environmental impact of AI, the campaign's creative approach also inspired participants' other works. P6 shared how they brought the campaign to other industry practitioners to reflect on how it could impact their work in the K-12 AI education curriculum. P4 reflected on their own research around satellite infrastructure, noting it gave them \textit{“ideas to toy with in terms of how do I talk about the impacts… in a way that is honestly quite similar to the struggles you may encounter with AI infrastructure.”} Similarly, P16 shared that the campaign inspired them to adapt the design concepts for their ongoing initiative on raising awareness on data harm.

As illustrated above, the various forms of resistance fostered by humour-based creative practices took shape as more approachable creative interventions, possibilities of new framings that de-stabilizes the dominant narrative, as well as the formation of a broader collective network of resistance that spans topics beyond, but related to, what is covered by our campaign. Our understanding of resistance aligns with the more nuanced framings of resistance from Sorøsen, Scott and Bayat as nonviolent, everyday resistance \cite{sorensen_humour_2016, scott_everyday_1989, bayat_life_2010}, that might not produce immediate political change, but instead operate through subtle acts embedded in everyday work and lives. As Bayata characterizes that these kind of resistance does not rely on protesting and demanding, but \textit{creating} alternative ways of thinking and acting \cite{bayat_life_2010}. For participants engaged with our campaign, resistance was also less about achieving immediate political outcomes, but more about influencing others through approachable actions, offering new ways of making sense of AI, and strengthening a network of practices that can work alongside each other toward reorienting the ways we think about and design technology. These forms of resistance stemming out of humourous creative practices occupy a distinctive role in resistance effort that are usually more sustainable \cite{sorensen_humour_2016}, and should not be ignored.

\section{Discussion}
\subsection{Humour-based Creative Practices for Sustainability Transformation}
\begin{figure*}
  \includegraphics[width=\textwidth]{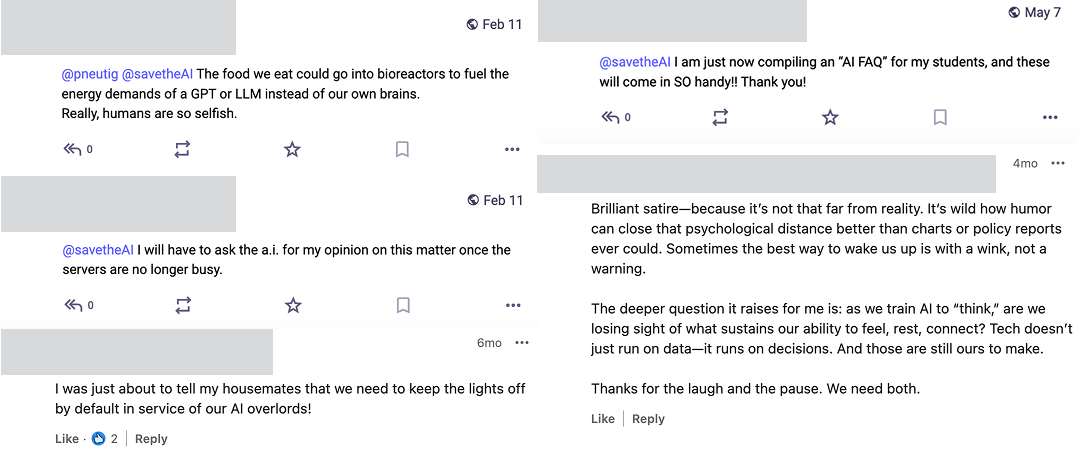}
  \caption{Examples of comments to re-posts of the campaign's digital content across social media platforms.}
  \label{fig:comments}
  \Description{A grid of five screenshots of our social media comments. The first one reads "The food we eat could go into bioreactors to fuel the energy demands of a GPT or LLM instead of our own brains. Really, humans are so selfish.". The second one reads "I will have to ask the ai. for my opinion on this matter once the servers are no longer busy.". The third one reads "I was just about to tell my housemates that we need to keep the lights off by default in service of our Al overlords!". The fourth one reads "I am just now compiling an "AI FAQ" for my students, and these will come in SO handy!! Thank you!" The fifth one reads "Brilliant satire-because it's not that far from reality. It's wild how humor can close that psychological distance better than charts or policy reports ever could. Sometimes the best way to wake us up is with a wink, not a warning. The deeper question it raises for me is: as we train Al to "think," are we losing sight of what sustains our ability to feel, rest, connect? Tech doesn't just run on data—it runs on decisions. And those are still ours to make. Thanks for the laugh and the pause. We need both."}
\end{figure*}

When initiating the campaign, we did not prioritize measuring quantitative metrics of engagement. However, at the time of writing this paper, we documented our posters displayed in 27 different cities across more than 10 countries, more than 1000 social media followers, thousands of interactions on our social media accounts, tens of thousands of post impressions, and numerous re-posts and discussions across social media platforms and local contexts (see figure \ref{fig:comments}) \cite{gestoso-souto_alert_2025, allhutter_international_2025, weston_dont_2025}. We also see our framings and jokes mentioned and echoed in articles, blog posts and online discussions \cite{white_jason_2025, simsek_rejection_2025, gilmour_no_2025}, which is exactly what we hoped for.

The campaign illustrates how humour-based creative practices can support efforts in response to long-standing calls to action in the field of sustainable HCI. In their review of a decade of sHCI research, Bremer et al. \cite{bremer_have_2022} synthesized recurring calls to action, including systems thinking, multidisciplinary collaboration, and supporting activism. By acting as a connector to critical friends and an emotional harbour, our humour-based creative campaign stimulated curiosity to \textit{learn}, made abstract data feel more \textit{embodied}, and brought together people who disagree in a space that allowed them to reflect and \textit{imagine} together. Under the CreaTures framework \cite{vervoort_9_2024}, learning, embodying, and imaging are key processes that stimulate \textcolor{NavyBlue}{\textbf{changes in meaning}}. In our study, such changes are illustrated through participants reflecting on designing smaller models (P1), questioning the effectiveness of eco-feedback tools (P2), and critically examining when generative AI tools should not be designed in the first place (P6, P8, P9, P11, P12). These reflections directly respond to calls in sHCI to move beyond simplified design “solutions,” attend to limits of design \cite{chen_strategy_2016, silberman_next_2014, laurell_thorslund_meta-crisis_2025, knowles_our_2022, hakansson_beyond_2013}, and when not to design at all \cite{baumer_when_2011}. 

Humour and creative practices also supports sHCI’s call for multidisciplinary collaboration \cite{bremer_have_2022, silberman_information_2015, chen_strategy_2016}. By acting as social glue, humour helps \textcolor{ForestGreen}{\textbf{change connections}} \cite{vervoort_9_2024}. Humour made it easier for us to connect with researchers in other disciplines as well as non-academic publics, leading to potential collaborations with community members and course development in disciplines such as American Studies (P15), Communications (P16, P17) and Education (P6). Humour and creative practices fostered a sense of inclusion and facilitated the formation of new relationships, enabling individuals to \textit{organize} together and \textit{inspire} one another to take action. 


By facilitating resistance, humour and creative works help provoke \textcolor{Maroon}{\textbf{changes in power}} \cite{vervoort_9_2024}, \textit{empowering} people to resist and to \textit{subvert} dominant narratives, and to \textit{co-create} new futures. Humour creates space for narratives within HCI that are usually pushed to the margins, offering their proponents agency and power. It also brings people together to co-create forms of collective resistance where they imagine, learn, empathise, define problems and act toward bringing new futures into being. The campaign responds to sHCI’s calls for supporting activism and mass movements \cite{laurell_thorslund_meta-crisis_2025, knowles_this_2018, qiao_near_2025}, contributing a tangible example, using creative practices to engage publics, inspire collective reflection, and open up opportunities to form coalitions with other activism work.

\subsection{Limits of Humour}
While humour plays important roles in facilitating collective reflection and resistance, it also has its limitations and risks. P5 reflected that: “\textit{Humour can't stand on its own to create justice because it doesn't legislate, and it doesn't regulate.}” There are also risks to using humour: humour can be considered unserious, it can be misunderstood by the audience, and even make people disillusioned \cite{sorensen_humour_2016}. For example, audiences interacting with the campaign may simply consume the materials \textit{“to have fun”} rather than engaging with the message behind the campaign. P3 mentioned that “\textit{[the poster designs] seemed like fun little things that could remind people of the impacts. But at the same time, it also [...] seemed like it could be tuned out in a way}”. The use of humour can sometimes be confusing or may lead to unintended perceived meanings: P6 and P11 mentioned not understanding the humour immediately and its connotations (i.e., whether the responsibility of the environmental and social impact of AI infrastructure lies with its users or with Big Tech). In this campaign, our website includes a clear statement explaining ``This should not be \textit{your} responsibility [...] Companies and policymakers must take responsibility''.

Beyond cognitive framings of humour \cite{martin_psychology_2007}, much of the literature recognizes humour as a culturally situated phenomenon \cite{kuipers_good_2006, apte_humor_1985, billig_laughter_2005}. What is perceived as playfully absurd and funny in one context may be seen as inappropriate and confusing in another context. In our campaign, we attempted to localize humour by having native speakers translating the posters using contextually relevant phrases to maintain the humourous part of the campaign. However, those who engaged more deeply with our study are often those who resonated with the humour, and this raises questions about gathering feedback from those who did not resonate with the campaign. Thus, future work is needed to understand the experience of those who might be alienated by the type of humour offered by our campaign, and to evaluate potential risks associated with humour-based interventions.

The subversive nature of this campaign's humour can also create risks. For instance, we heard from several colleagues in academia and in industry who felt strongly, positively, about the campaign but felt they had to refrain from putting up posters in their workplaces for fear of being `found out' as their originators and facing negative consequences for the criticism in their workplaces. Both understandable and unfortunate, it suggests that in challenging tech power, the campaign is hitting a nerve.

\subsection{Future Work Beyond Humour}
Our campaign primarily poked fun at the absurd amount of natural and social resources consumed by AI infrastructures, but we recognize that a wide variety of creative practices (i.e. creative placemaking \cite{miller_view_2022}, immersive role play game \cite{furtherfield_treaty_2022}, sculpture installation \cite{beavers_nocturne_2022}) are actively working toward “eco-social transformations” \cite{creatures_creatures_2022} and we especially see creative practices’ power in helping people to imagine alternative futures in an embodied way \cite{houston_creatures_2022}. In future work, we seek to collaborate with artists and engage in a wider range of creative practices.


Following the call for transformative pedagogy \cite{houston_creatures_2022}, we see future creative endeavors particularly well suited to teaching and research contexts. We are developing workshops that can be embedded in longer term courses. Inspired by our participants (P6, P7, P16, P17), we are building collaborations with educators across disciplines such as philosophy and sustainability pedagogy. We also see value in conducting longitudinal study to examine how sustained engagement in creative practices shapes students’ relationships to socio-ecological transformation over time. We are also keen to work with activists to develop forms of resistance to data centre expansion by incorporating creative practices and exploring their potential for systems change.

\section{Conclusion}
Commentators on our early posts about the ``thirst'' of AI wondered, ``how much time would it take for this to become real''. Now we know: within several months, headlines reported that ``AI Data Centers in Texas Used 463 Million Gallons of Water, Residents Told to Take Shorter Showers” \cite{anthony_ai_2025} and ``Their Water Taps Ran Dry When Meta Built Next Door'' \cite{tan_their_2025}. These were just two of many stories illustrating how the speculated dystopia in our satire are becoming reality, underscoring the urgency of sustaining momentum and expanding efforts to draw attention to the environmental and social impacts of AI infrastructure. Encouragingly, one year after our launch, we begin to see many other humorous and creative campaigns \cite{quiliai_quiliai_2026, caaac_center_2025, replacementai_replacementai_2025} making visible these concerns, as well as large scale resistance to data centres leading to moratoriums \cite{national_conference_of_state_legislatures_which_2026, roach_denmark_2026, data_center_bans_ai_2026}.

In this paper, we reflected on humour-based creative practices' roles in addressing the environmental and social impact of AI infrastructure. By designing artifacts, facilitating workshops, and engaging participants through surveys and interviews, we showed how humour and creative work made abstract technological concerns tangible and fostered inclusive dialogue highlighting perspectives often overlooked in discussions around AI. 

As a connector to critical friends, humour surfaces uncomfortable topics in approachable ways that prompt people to reflect and learn; as a visceral and emotional harbour, humour opens up an affective space for people to have embodied and emotional responses to otherwise distant data; as social glue, humour fosters inclusion and breaks down barriers for people to connect with each other; as resistance, humour undermines dominant narratives and empowers sustained collective actions. For HCI, our work highlights humour-based creative practices not simply as communication tools, but as methods for collective reflection and critique that expands how we approach societal and ecological implications of emerging technologies. Humour and creative practices, when mobilized thoughtfully, inspire new forms of awareness, solidarity, and resistance in pursuing sustainability transformations.

\begin{acks}
Beyond the core research team who are listed as the authors of this paper, many people contributed to the campaign and supported us along the way. We hope to acknowledge their contribution to the brainstorming, design, and installation processes with immense gratitude: Nils Bonfils, Monica Iqbal, Olya Jaworsky, Victoria GD Landau, Lena Rubisova, Ayusha Thapa, Cristian Velasquez, and friends of the Just Sustainability Design Lab. Many thanks to our friends, colleagues and supporters who helped with translation, and placed posters in cities around the globe. The research project was partially supported by NSERC Discovery Grant RGPIN-2025-07063.
\end{acks}

\bibliographystyle{ACM-Reference-Format}
\bibliography{SavetheAI,EB,RM}


\end{document}